\documentclass[11pt,fleqn]{article} 
\usepackage{graphicx}
\usepackage{textcomp} 
\usepackage{mathtools}
\usepackage{epstopdf}
\usepackage{latexsym}
\usepackage{amsmath}
\usepackage{amssymb}
\usepackage{bm}
\usepackage[mathcal]{euscript}
\usepackage{slashed}
\usepackage{tikz}
\usepackage{MnSymbol}
\usepackage{cancel}

\usepackage[top=30mm, bottom=30mm, left=25mm, right=25mm]{geometry}

\numberwithin{equation}{section}

\usepackage{marvosym}

\usepackage{wasysym}

\usepackage{indentfirst}

\newcommand\blfootnote[1]{
  \begingroup
  \renewcommand\thefootnote{}\footnote{#1}
  \addtocounter{footnote}{-1}
  \endgroup
}

\usepackage{hyperref}
\hypersetup{
        unicode,	
        colorlinks,
        citecolor=blue,
        linkcolor=blue, 
        urlcolor=blue,
        bookmarksopen=true,
        bookmarksopenlevel=\maxdimen,
      }

\def\gl#1#2{\ifmmode \mathrm{GL}(#1; {\bf #2}) \else $\mathrm{GL}(#1; {\bf #2})$\fi}
\def\sl#1#2{\ifmmode \mathrm{SL}(#1; {\bf #2}) \else $\mathrm{SL}(#1; {\bf #2})$\fi}
\def\so#1{\ifmmode \mathrm{SO}({#1}) \else $\mathrm{SO}(#1)$\fi}

\def\sp#1#2{\ifmmode \mathrm{Sp}(#1; {\bf #2}) \else $\mathrm{Sp}(#1; {\bf #2})$\fi}
\def\usp#1{\ifmmode \mathrm{USp}(#1) \else $\mathrm{USp}(#1)$\fi}
\def\spin#1{\ifmmode \mathrm{Spin}(#1) \else $\mathrm{Spin}(#1)$\fi}
\def\su#1{\ifmmode \mathrm{SU}({#1}) \else $\mathrm{SU}(#1)$\fi}

\def\double #1{#1{\hbox{\kern-2pt $#1$}}}

\mathcode`\*="702A                  
\def\half{{\textstyle{1\over{\raise.1ex\hbox{$\scriptstyle{2}$}}}}}

\def \p{\partial}

\def \d{\delta}

\def \L{\Lambda}

\def \s{\sigma}

\def\bt{\widetilde b}
\def\ct{\widetilde c}

\begin{document}

\begin{flushright}
\makebox[0pt][b]{}
\end{flushright}

\hspace{12cm} 

\vspace{40pt}
\begin{center}
{\LARGE \bf{A note on gauge-covariant vertex operators of the ambitwistor string}}



\vspace{40pt}
Osvaldo Chandia${}^{\star}$
\vspace{20pt}

{\em Departamento de Ciencias, Facultad de Artes Liberales
}\\
{\em Universidad Adolfo Ib\'a\~nez, Chile}\\



\vspace{60pt}
{\bf Abstract}
\end{center}
We construct gauge-covariant unintegrated vertex operators in an
antighost-independent subcomplex of the bosonic ambitwistor string.  For a
linearized gauge field, a world-sheet ghost-derivative term completes the
standard vertex so that BRST closure imposes
$\p^nF_{nmI}=0$ without the Lorenz condition.  In the rank-two sector we use
an antighost-free ansatz that is sufficient to isolate a diffeomorphism-invariant
spin-two subsector.  After eliminating auxiliary fields and consistently
setting the antisymmetric and scalar combinations to zero, BRST closure gives
$R_{mn}=0$ without imposing transversality.  We distinguish this
construction from the full relative BRST cohomology of the bosonic
ambitwistor string, whose additional antighost-dependent states lead to the
known higher-derivative spectrum.

\blfootnote{
${}^{\star}$ \href{mailto:ochandiaq@gmail.com}{ochandiaq@gmail.com}}


\setcounter{page}0
\thispagestyle{empty}
\newpage

\tableofcontents

\parskip = 0.1in

\section{Introduction}
Ambitwistor strings are chiral world-sheet theories with only massless
space-time states \cite{Mason:2013sva}. Their BRST cohomology relates
unintegrated vertex operators to linearized equations of motion and gauge
transformations. The simplest representatives, however, often impose a
space-time gauge condition in addition to the field equation.

Gauge-covariant representatives avoid this feature by including derivatives
of the world-sheet ghosts.  This mechanism was studied for conventional
strings in \cite{Siegel:2003sv} (see also \cite{Chandia:2021coc}). Here we apply it to a
restricted, antighost-independent sector of the bosonic ambitwistor string.

It is important to state the scope of the calculation. The full relative
ghost-number-two cohomology contains antighost-dependent operators and has a
higher-derivative, non-unitary bosonic spectrum \cite{Berkovits:2018jvm}.  Related
ambitwistor descriptions of $R^2$ gravity and $(DF)^2$ gauge theory were
developed in \cite{Azevedo:2017lkz,Azevedo:2019zbn}, and the full bosonic
spectrum has recently been revisited in \cite{Figueroa-OFarrill:2026igk}.  We do not
recompute or replace that spectrum.  Instead, we identify explicit
gauge-covariant BRST representatives in a smaller subcomplex and show that it
contains a consistent linearized Einstein subsector.  This distinction also
means that the gauge-field result below is a statement about the free
equation; it does not identify the interacting theory with ordinary
Yang-Mills theory.

Section~\ref{sec:model} defines the model and the restricted complex.
Section~\ref{sec:ym} treats the gauge field.  Section~\ref{sec:grav} derives
the spin-two representative and proves the consistency of the two
truncations.  Section~\ref{sec:relation} compares the result with the full
cohomology. Calculations are shown in the appendices.

\section{The model and the restricted operator complex}
\label{sec:model}

The gauge-fixed chiral action is
\begin{align}\label{action}
 S=\int d^2z\left(P_m\bar\p X^m+
 \frac12\rho_A\bar\p\rho_A+b\bar\p c+\bt\bar\p\ct\right),
\end{align}
where $(X^m,P_m)$ have conformal weights $(0,1)$, while $(b,c)$ and
$(\bt,\ct)$ have weights $(2,-1)$.  The real fermions $\rho_A$ generate the
current algebra used in Section~\ref{sec:ym}.  The BRST charge is
\begin{align}\label{Q}
 Q=\oint\frac{dz}{2\pi i}\left[cT+\ct H+bc\p c+
 \bt(c\p\ct+\ct\p c)\right],
\end{align}
with $T=-P_m\p X^m-\frac12\rho_A\p\rho_A, H=-\frac12P_mP^m$ in the matter sector.
Nilpotence requires
\begin{equation}
 D=26-\frac{N}{4},
\end{equation}
where $N$ is the number of real current-algebra fermions.  When we say below
that a gravity vertex has ``vanishing fermions,'' we mean that the operator is
a current-algebra singlet and contains no explicit $\rho_A$; the fermions
remain in the world-sheet conformal field theory and continue to contribute
to this criticality condition.

Let $C_0$ denote the graded space of local operators built from
$X,P,\rho,c,\ct$ and their derivatives but containing neither $b$ nor $\bt$.
Directly from \eqref{Q}, $Q C_0\subset C_0$, so it is a BRST
subcomplex.  We study its ghost-number-two, conformal-weight-zero component,
and restrict gauge parameters in the same way. Conformal weight zero is the
local-operator form of the $L_0=0$ condition.  We do not claim that
$C_0$ is the most general relative complex: in particular,
the $b_0$-relative analysis of \cite{Berkovits:2018jvm} admits additional
antighost-dependent combinations.  The final representatives constructed
below are themselves $b_0$-closed, but their possible equivalence to states
outside $C_0$ is not addressed here.

Our conventions are
$A_{(mn)}=A_{mn}+A_{nm}$ and $A_{[mn]}=A_{mn}-A_{nm}$; thus symmetrization and
antisymmetrization carry no factor of $1/2$.  A world-sheet derivative is
written as $\p$, while $\p_m$ is a space-time derivative.

\section{Gauge-covariant linearized gauge field}
\label{sec:ym}

Let define 
\begin{align}
 J^I=\frac12K^I{}_{AB}\rho_A\rho_B,
\end{align}
where $K^I{}_{AB}=-K^I{}_{BA}$ are generators in the representation of
$\rho_A$. At the linearized level, only the conformal weight and the adjoint
index of $J^I$ enter the calculation. Define
$F_{mnI}=\p_{[m}A_{n]I}$. The gauge-covariant vertex is
\begin{align}
 U_A=c\ct P^mJ^IA_{mI}(X)
      -\frac12c\p\ct J^I\p^mA_{mI}(X).
 \label{UYM}
\end{align}
Using the OPEs in Appendix~\ref{app:ope}, one finds
\begin{align}
 QU_A=-\frac12c\ct\p\ct P^mJ^I\p^nF_{nmI}.
\end{align}
Therefore, $QU_A=0$ gives the gauge-covariant free equation
\begin{align}
 \p^nF_{nmI}=0,
\end{align}
without imposing $\p^mA_{mI}=0$.  Moreover,
\begin{align}
 \delta U_A=Q\left(-cJ^I\Lambda_I(X)\right),
\end{align}
induces $\delta A_{mI}=\p_m\Lambda_I$.  This construction concerns the
linearized equation only.  Ambitwistor models with current algebra have a
known higher-derivative interpretation $(DF)^2$ at the interacting level
\cite{Azevedo:2017lkz,Azevedo:2019zbn}.

\section{Gauge-covariant spin-two representative}
\label{sec:grav}

The conventional vertex
$c\ct P^mP^nA_{mn}$ is BRST closed only after imposing both a wave equation
and transversality.  We instead seek the covariant equation in which the linearized Ricci tensor vanishes, that is,
\begin{align}
R_{mn}=-\frac12\left(
 \Box A_{mn}-\p_{(m}\p^pA_{n)p}
 +\p_m\p_n A^p{}_p\right)=0.
\label{Ricci}
\end{align}
We begin with the following antighost-free weight-zero ansatz in
$C_0$:
\begin{align}
U&=c\ct P^mP^nA_{mn}+c\ct P^m\p X^nB_{mn}
 +c\ct\p X^m\p X^nC_{mn}+c\ct\p P^mD_m+c\ct\p^2X^mE_m
\cr
&+c\p\ct P^mF_m+c\p\ct\p X^mG_m
 +\ct\p cP^mH_m+\ct\p c\p X^mI_m
 +\p c\p\ct J+c\p^2\ct K+\ct\p^2cL
\cr
&+c\p cP^mM_m+c\p c\p X^mN_m
 +\ct\p\ct P^m\widetilde M_m+\ct\p\ct\p X^m\widetilde N_m
 +c\p^2cO+\ct\p^2\ct\widetilde O.
 \label{generalU}
\end{align}
This is a sufficient ansatz for the representative sought here, not the most
general vertex of the full relative cohomology.  The corresponding restricted
gauge parameter is
\begin{align}
 \Lambda=cP^m\lambda_m+c\p X^m\rho_m+
 \ct P^m\widetilde\lambda_m+
 \ct\p X^m\widetilde\rho_m+\p c\,\sigma+\p\ct\,\widetilde\sigma.
 \label{gaugeparameter}
\end{align}

The algebraic gauge transformations can eliminate
$H_m,E_m,L,D_m$ together with their dependent partners. The detailed calculation is performed in appendix \ref{sec3}. BRST closure then organizes the remaining fields into a $A$-sector and a homogeneous $B$-sector. For the latter one obtains
\begin{align}\label{Bsector}
&K=-\frac16\eta^{mn}B_{mn},\quad G_m=-\frac12\p^nB_{nm},\cr 
&\p^nB_{(mn)}-\frac13\p_m(\eta^{np}B_{np})=0,\cr 
&C_{mn}=\frac14\Box B_{mn}-\frac14\p_{[m}\p^pB_{pn]},\cr 
&\widetilde O=\frac16\eta^{mn}C_{mn}
=\frac14\p^m\p^nB_{mn}.
\end{align}
All these equations are homogeneous in $B_{mn}$. Hence
\begin{equation}
 B_{mn}=0
 \label{Bzero}
\end{equation}
consistently sets $C_{mn},G_m,K,\widetilde O$ to zero. Note that this is a truncation, not a gauge choice. It is preserved by the diffeomorphism parameter
$\Lambda_\lambda=cP^m\lambda_m$, because the transformation of $B_{mn}$
depends only on $\widetilde\lambda_m$ and $\rho_m$, both of which are set to
zero in this subsector.

After \eqref{Bzero}, the nonzero part of $QU$ reduces to
\begin{align}\label{QUreduced}
QU&=c\p^2c\ct P^m(\p^nA_{mn}+F_m+\p_mO)
\nonumber\\
&+\frac12c\ct\p\ct P^mP^n(-\Box A_{mn}-\p_mF_n)
 +\frac12c\p^2c\p\ct(\p^mF_m-\Box O).
 \end{align}
The first coefficient gives
\begin{align}
 F_m=-\p^n A_{mn}-\p_m O.
 \label{Fsolution}
\end{align}
Since $P^m P^n$ is symmetric, the symmetric part of the second coefficient
then gives
\begin{align}
 \Box A_{mn}-\p_{(m}\p^pA_{n)p}-2\p_m\p_nO=0.
 \label{AeqO}
\end{align}

Under $\Lambda_\lambda$,
\begin{align}
 \delta A_{mn}=-\frac12\p_{(m}\lambda_{n)},
 \quad
 \delta O=\frac12\p^m\lambda_m.
 \label{diffs}
\end{align}
Consequently, the scalar combination
\begin{align}
 \Phi=A^m{}_m+2O
 \label{Phi}
\end{align}
is gauge invariant.  The pure-spin-two restriction is the gauge-invariant
truncation $\Phi=0$, or
\begin{align}
 O=-\frac12A^m{}_m.
 \label{Osolution}
\end{align}
Equation~\eqref{AeqO} now becomes
\begin{align}
 \Box A_{mn}-\p_{(m}\p^pA_{n)p}
 +\p_m\p_nA^p{}_p=0,
\end{align}
which is equivalent to $R_{mn}=0$.  The last coefficient in
\eqref{QUreduced} is not independent, since
\begin{align}
 \p^m F_m-\Box O
 =\Box A^m{}_m-\p^m\p^nA_{mn}
 =-\eta^{mn}R_{mn}
\end{align}
in the conventions of \eqref{Ricci}.

The resulting vertex is
\begin{align}
U_g=c\ct P^mP^nA_{mn}
 -c\p\ct P^m\left(\p^nA_{mn}-\frac12\p_mA^n{}_n\right)
 -\frac12c\p^2c\,A^m{}_m .
 \label{finalvertex}
\end{align}
It obeys
\begin{align}
 \delta U_g=Q(cP^m\lambda_m),
\end{align}
with the diffeomorphism in \eqref{diffs}, and $QU_g=0$ is equivalent to
$R_{mn}=0$.

Within the restricted complex, the class is nontrivial for ordinary
spin-two-plane waves.  An exact deformation generated by
$cP^m\lambda_m$ is a pure linearized diffeomorphism and therefore has
a vanishing linearized Riemann tensor. A solution with nonzero linearized Weyl
curvature cannot be of this form.  This establishes nontriviality in the
spin-two subcomplex; it is not a new classification of the complete bosonic
ambitwistor cohomology.

\section{Relation to the full bosonic cohomology}
\label{sec:relation}

The distinction between \eqref{generalU} and the general relative vertex is
essential.  The latter contains, among other structures, terms of the form
$\bt\ct c\p\ct\,S$ and $bc\p\ct\ct\,S'$ as well as corresponding
antighost-dependent gauge parameters \cite{Berkovits:2018jvm}.  Retaining those
operators leads to additional metric, antisymmetric-tensor, scalar and vector
sectors and to a higher-derivative kinetic action.  Curved-background vertex
operators likewise naturally treat the graviton, two-form and dilaton
together \cite{Adamo:2018ege}.

Our construction answers a different and narrower question: whether one can
complete the familiar $c\ct P^mP^nA_{mn}$ operator, using only derivatives of
the positive-ghost-number fields, so that BRST closure is gauge covariant
rather than transverse-gauge fixed.  Equation~\eqref{finalvertex} gives such
a completion. The homogeneous $B$ equations and the invariant scalar
$\Phi$ show explicitly how the Einstein branch is isolated. No conclusion
about the interacting completion of this branch is implied.

\section{Conclusion}

We constructed gauge-covariant BRST representatives in an explicitly defined
antighost-independent subcomplex of the bosonic ambitwistor string.  In the
current-algebra sector, a ghost-derivative completion converts the usual
transverse representative into one obeying the covariant linearized gauge
equation.  In the rank-two sector, $B_{mn}=0$ is a homogeneous truncation and
$\Phi=A^m{}_m+2O=0$ is a gauge-invariant scalar truncation. On their
intersection BRST closure of \eqref{finalvertex} is precisely the
linearized Einstein equation, while BRST-exact deformations generate
linearized diffeomorphisms.  The construction isolates an Einstein
subcomplex and is complementary to, rather than a replacement for, the known
higher-derivative full cohomology.

\paragraph{AI disclosure.}
During the preparation of this work, the author used generative AI tools for
structural organization and language refinement.  The author takes full
responsibility for the manuscript's ideas, calculations and conclusions.

\appendix
\section{OPE conventions}
\label{app:ope}

All composite operators below are normal ordered. The elementary free-field
OPEs are
\begin{equation}
 P_m(y)X^n(z)\to-\frac{\delta_m^n}{y-z},\quad
 b(y)c(z)\to\frac1{y-z},\quad
 \bt(y)\ct(z)\to\frac1{y-z},\quad
 \rho_A(y)\rho_B(z)\sim\frac{\delta_{AB}}{y-z}.
\end{equation}
From here the needed OPEs are
\begin{align}
    T(y)P^m P^n f_{mn}(z)&\to-\frac{2}{(y-z)^3}P^m\p^n f_{mn}(z)+\frac{2}{(y-z)^2}P^m P^n f_{mn}(z)\cr 
    &+\frac{1}{(y-z)}\p(P^m P^n f_{mn}(z)),
\end{align}
\begin{align}
    T(y) P^m\p X^n f_{mn}(z)&\to-\frac{1}{(y-z)^4}\eta^{mn}f_{mn}(z)-\frac{1}{(y-z)^3}\p X^m\p^n f_{nm}(z)\cr&+\frac{2}{(y-z)^2}P^m \p X^n f_{mn}(z)+\frac{1}{(y-z)}\p(P^m \p X^n f_{mn}(z)),
\end{align}
\begin{align}
    T(y)\p X^m\p X^n f_{mn}(z)\to\frac{2}{(y-z)^2} \p X^m\p X^n f_{mn}(z)+\frac{1}{(y-z)} \p(\p X^m\p X^n f_{mn}(z)),
\end{align}
\begin{align}
    T(y)\p P^m f_m(z)&\to-\frac{2}{(y-z)^4}\p^m f_m(z)+\frac{2}{(y-z)^3}P^m f_m(z)+\frac{2}{(y-z)^2}\p P^m f_m(z)\cr&+\frac{1}{(y-z)}\p(\p P^m f_m(z)),
\end{align}
\begin{align}
    T(y)\p^2X^m f_m(z)&\to\frac{2}{(y-z)^3}\p X^m f_m(z)+\frac{2}{(y-z)^2}\p^2X^m f_m(z)\cr
    &+\frac{1}{(y-z)}\p(\p^2X^m f_m(z)),
\end{align}
\begin{align}
    T(y) P^mf_m(z)&\to-\frac{1}{(y-z)^3}\p^m f_m(z)+\frac{1}{(y-z)^2}P^mf_m(z)+\frac{1}{(y-z)}\p(P^mf_m(z)),
\end{align}
\begin{align}
    T(y)\p X^m f_m(z)\to\frac{1}{(y-z)^2}\p X^m f_m(z)+\frac{1}{(y-z)}\p(\p X^m f_m(z)),
\end{align}
\begin{align}
    T(y) f(z)\to\frac{1}{(y-z)}\p f(z),
\end{align}
\begin{align}
    H(y)f(z)\to-\frac{1}{(y-z)^2}\frac12\Box f(z)+\frac{1}{(y-z)} P^m \p_m f(z),
\end{align}
\begin{align}
    H(y)\p X^m f_m(z)&\to-\frac{1}{(y-z)^3}\p^m f_m(z)+\frac{1}{(y-z)^2}(P^m f_m(z)-\frac12\p X^m\Box f_m(z))\cr 
    &+\frac{1}{(y-z)}(\p P^m f_m(z)+P^m\p X^n \p_m f_n(z)),
\end{align}
\begin{align}
    H(y)\p X^m\p X^n f_{mn}(z)&\to-\frac{1}{(y-z)^4}\eta^{mn}f_{mn}(z)-\frac{2}{(y-z)^3}\p X^m\p^n f_{mn}(z)\cr 
    &+\frac{1}{(y-z)^2}(2P^m\p X^n f_{mn}(z)-\frac12\p X^m\p X^n\Box f_{mn}(z))\cr 
    &+\frac{1}{(y-z)}(2\p P^m \p X^n f_{mn}(z)+P^p\p X^m\p X^n\p_pf_{mn}(z)),
\end{align}
\begin{align}
    H(y)\p^2 X^m f_m(z)&\to-\frac{2}{(y-z)^4}\p^m f_m(z)+\frac{2}{(y-z)^3}P^m f_m(z)\cr 
    &+\frac{1}{(y-z)^2}(2\p P^m f_m(z)-\frac12\p^2 X^m\Box f_m(z))\cr 
    &+\frac{1}{(y-z)}(\p^2 P^m f_m(z)+P^m\p^2 X^n \p_m f_n(z)) ,
\end{align}

\begin{align}\label{need}
    &T(y)P^m J^I f_{mI}(z)\to-\frac{1}{(y-z)^3}J^I \p^m f_{mI}(z)+\frac{2}{(y-z)^2}P^m J^I f_{mI}(z)+\frac{1}{(y-z)}\p(P^m J^I f_{mI}(z)),\cr
    &T(y) J^I f_I(z)\to\frac{1}{(y-z)^2}J^I f_I(z)+\frac{1}{(y-z)}\p(J^I f_I(z)),
\end{align}
where $f's$ are functions of $X$.

\section{Construction of $U$}\label{sec3}
Consider the most general conformal weight zero and ghost number $+2$ vertex. It has the form
\begin{align}\label{genU}
    U&=c\ct P^m P^n A_{mn}(X)+c\ct P^m \p X^n B_{mn}(X)+c\ct\p X^m\p X^n C_{mn}(X)+c\ct\p P^m D_m(X)\cr
    &+c\ct\p^2 X^m E_m(X)+c\p\ct P^m F_m(X)+c\p\ct\p X^m G_m(X)+\ct\p c P^m H_m(X)\cr
    &+\ct\p c\p X^m I_m(X)+\p c\p\ct J(X)+c\p^2\ct K(X)+\ct\p^2 c L(X)+c\p c P^m M_m(X)\cr 
    &+c\p c \p X^m N_m(X)+\ct\p\ct P^m {\widetilde M}_m(X)+\ct\p\ct\p X^m{\widetilde N}_m(X)+c\p^2 c O(X)+\ct\p^2\ct {\widetilde O}(X).
\end{align}
This vertex is defined up to $Q\L$. A general ghost number $+1$ and conformal weight zero gauge parameter $\L$ has the form 
\begin{align}\label{genLambda}
\Lambda={}
cP^m\lambda_m(X)
+c\p X^m\rho_m(X)
+\ct P^m\widetilde\lambda_m(X)
+\ct\p X^m\widetilde\rho_m(X)
+\p c\sigma(X)
+\p\ct\widetilde\sigma(X).
\end{align}
$\d U=Q\L$ implies,
\begin{align}\label{dU}
&\delta A_{mn}=-\frac12\p_{(m}\lambda_{n)},\quad \delta B_{mn}=\p_n\widetilde\lambda_m-\p_m\rho_n,
\quad
\delta C_{mn}=\frac12\p_{(m}\widetilde\rho_{n)},\quad \delta D_m=\widetilde\lambda_m-\rho_m,\cr 
&\delta E_m=\widetilde\rho_m,
\quad
\delta F_m=\frac12\Box\lambda_m-\rho_m+\widetilde\lambda_m,
\quad
\delta G_m=\frac12\Box\rho_m+\widetilde\rho_m
+\p_m\widetilde\sigma,\quad \delta H_m=\p_m\sigma,
\cr 
&\delta I_m=0,\quad
\delta J=\frac12\Box\sigma,
\quad
\delta K=\frac12\p^m\rho_m+\widetilde\sigma,
\quad
\delta L=\frac12\p^m\widetilde\lambda_m+\widetilde\sigma,
\quad
\delta M_m=0,
\quad
\delta N_m=\p_m\sigma,
\cr
&\delta\widetilde M_m=\frac12\Box\widetilde\lambda_m
-\widetilde\rho_m+\p_m\widetilde\sigma,
\quad
\delta\widetilde N_m=\frac12\Box\widetilde\rho_m,
\quad
\delta O=\frac12\p^m\lambda_m+\sigma,
\quad
\delta\widetilde O=\frac12\p^m\widetilde\rho_m.
\end{align}
Before gauge fix some of the fields using these transformations let's find the equations from $QU=0$. Using the form of $Q$ \eqref{Q} and the OPEs of the appendix we obtain
\begin{align}\label{QU=0}
    &QU=c\p^3c\ct\left(\frac16\eta^{mn}B_{mn}+\frac13\p^m D_m+K-L\right)+c\ct\p^3\ct\left(-\frac16\eta^{mn} C_{mn}-\frac13\p^m E_m+\widetilde O\right)\cr 
    &+c\p^2c\ct P^m\left(\p^n A_{mn}-D_m+F_m-H_m+\p_m O\right)+c\p^2c\ct\p X^m\left(\frac12\p^n B_{nm}-E_m+G_m-I_m-\p_m L\right)\cr 
    &+c\ct\p^2\ct P^m\left(-\frac12\p^n B_{mn}+E_m-\p_m K+{\widetilde M}_m\right)+c\ct\p^2\ct\p X^m\left(-\p^n C_{mn}+{\widetilde N}_m+\p_m{\widetilde O}\right)\cr 
    &+c\ct\p\ct P^m P^n\left(-\frac12\Box A_{mn}+B_{mn}-\p_m F_n\right)+c\ct\p\ct P^m\p X^n\left(-\frac12\Box B_{mn}+2C_{mn}-\p_m G_n+\p_n{\widetilde M}_m\right)\cr 
    &+c\ct\p\ct\p X^m\p X^n\left(-\frac12\Box C_{mn}+\p_m{\widetilde N}_m\right)+c\ct\p\ct\p P^m\left(-\frac12\Box D_m+2E_m -G_m+{\widetilde M}_m\right)\cr 
    &+c\ct\p\ct\p^2 X^m\left(-\frac12\Box E_m+{\widetilde N}_m\right)+c\p c\ct P^m P^n\left(\p_m M_n\right)+c\p c\ct\p P^m\left(-H_m+N_m\right)\cr 
    &+c\p c\ct P^m\p X^n\left(-\p_n H_m+\p_m N_n\right)+c\p c\ct\p^2 X^m\left(-I_m\right)+c\p c\ct\p X^m\p X^n\left(-\p_m I_n\right)\cr 
    &+c\p c\p\ct P^m\left(-H_m+N_m-\frac12\Box M_m\right)+c\p c\p\ct\p X^m\left(-I_m+\p_m J-\frac12\Box N_m\right)\cr 
    &+\p c\ct\p\ct P^m\left(\frac12\Box H_m-I_m-\p_m J\right)+\p c\ct\p\ct \p X^m\left(\frac12\Box I_m\right)+c\p^2 c\p\ct\left(\frac12\p^m F_m+J+K-L-\frac12\Box O\right)\cr 
    &+c\p\ct\p^2\ct\left(-\frac12\p^m G_m+\frac12\Box K+{\widetilde O}\right)+\p c\p^2 c\ct\left(-\frac12\p^m H_m+J\right)+\p c\ct\p^2\ct\left(\frac12\p^m I_m\right)\cr
    &+c\p c\p^2\ct\left(J-\frac12\p^m N_m\right)+c\p c\p^2 c\left(-\frac12\p^m M_m\right)+\p^2 c\ct\p\ct\left(\frac12\Box L-\frac12\p^m{\widetilde M}_m-{\widetilde O}\right)\cr 
    &+\ct\p\ct\p^2\ct\left(-\frac12\p^m{\widetilde N}_m+\frac12\Box{\widetilde O}\right),
\end{align}
and impose the equation $QU=0$. The vanishing of the term involving $c\p c\ct\p^2 X^m$ in the seventh line of \eqref{QU=0} implies $I_m=0$. Similarly, the vanishing of the term with the factor of $c\p c\ct\p P^m$ in the sixth line of \eqref{QU=0} implies $N_m=H_m$. Vanishing the term with $c\p c\ct P^m\p X^m$ in the seventh line of \eqref{QU=0} implies
\begin{align}\label{eqH}
    \p_{[m}H_{n]}=0,
\end{align}
which up to topological obstructions that we assume absent implies that 
\begin{align}\label{His}
    H_m=\p_m h,
\end{align}
for a scalar function $h(x)$. Recalling the gauge invariance for $H_m$ in \eqref{dU} is $\d U=\p_m\s$, we can gauge fix $H_m=0$. Consequently, $N_m$ and $J$ also vanish. We now use the gauge invariance $\d E_m={\widetilde\rho}_m$ from \eqref{dU} to gauge fix $E_m=0$ and the vanishing of the term with the factor $c\ct\p\ct\p^2 X^m$ in the sixth line of \eqref{QU=0} implies ${\widetilde N}_m=0$. Using now the gauge invariance $\d L=\frac12\p^m{\widetilde\lambda}_m+{\widetilde\sigma}$ to fix $L=0$ and finally, the gauge invariance $\d D_m={\widetilde\lambda}_m-\rho_m$ allows to fix $D_m=0$. The vanishing of the term with the factor $c\ct\p\ct\p P^m$ in the fifth line of \eqref{QU=0} implies ${\widetilde M}_m=G_m$. Up to now we have
\begin{align}\label{QUmid}
    &QU=c\p^3c\ct\left(\frac16\eta^{mn}B_{mn}+K
    \right)+c\ct\p^3\ct\left(-\frac16\eta^{mn} C_{mn}+\widetilde O\right)\cr 
    &+c\p^2c\ct P^m\left(\p^n A_{mn}+F_m+\p_m O\right)+c\p^2c\ct\p X^m\left(\frac12\p^n B_{nm}+G_m\right)\cr 
    &+c\ct\p^2\ct P^m\left(-\frac12\p^n B_{mn}-\p_m K+G_m\right)+c\ct\p^2\ct\p X^m\left(-\p^n C_{mn}+\p_m{\widetilde O}\right)\cr 
    &+c\ct\p\ct P^m P^n\left(-\frac12\Box A_{mn}+B_{mn}-\p_m F_n\right)+c\ct\p\ct P^m\p X^n\left(-\frac12\Box B_{mn}+2C_{mn}-\p_m G_n+\p_n G_m\right)\cr 
    &+c\ct\p\ct\p X^m\p X^n\left(-\frac12\Box C_{mn}\right)+c\p^2 c\p\ct\left(\frac12\p^m F_m+K-\frac12\Box O\right)+\ct\p\ct\p^2\ct\left(-\frac12\Box{\widetilde O}\right)\cr 
    &+c\p\ct\p^2\ct\left(-\frac12\p^m G_m+\frac12\Box K+{\widetilde O}\right)+\p^2 c\ct\p\ct\left(-\frac12\p^m G_m-{\widetilde O}\right)\cr
    &+c\p c\ct P^m P^n\left(\p_m M_n\right)+c\p c\p\ct P^m\left(-\frac12\Box M_m\right)+c\p c\p^2 c\left(-\frac12\p^m M_m\right)=0.
\end{align}
Note that $M_m$ appears in the last line of \eqref{QUmid} is decoupled so we ignore this field from now on. 

The fields $G_m, K$ are functions of $B$. In fact, the vanishing of term with the factor of $c\p^3 c\ct$, and the vanishing of term with the factor of $c\p^2 c\ct\p X^m$ in \eqref{QUmid} give
\begin{align}\label{KGm}
    K=-\frac16\eta^{mn} B_{mn},\quad G_m=-\frac12\p^n B_{nm}. 
\end{align}
Given this result, we find an equation for $B_{mn}$ from vanishing the term with $c\ct\p^2\ct P^m$ in \eqref{QUmid}. It gives
\begin{align}\label{eqB}
    \p^n B_{(mn)}-\frac13\p_m\left(\eta^{np}B_{np}\right)=0.
\end{align}
Acting with $\p^m$ here one obtains
\begin{align}\label{eqBB}
    \p^m\p^n B_{mn}=\frac16\Box\left(\eta^{mn} B_{mn}\right).
\end{align}
Similarly, the vanishing of the term with the factor of $c\ct\p\ct P^m\p X^n$ in \eqref{QUmid} determines $C_{mn}$ as
\begin{align}\label{Cis}
    C_{mn}=\frac14\Box B_{mn}-\frac14\p_{[m}\p^p B_{pn]},
\end{align}
and vanishing the term with the factor $c\ct\p^3\ct$ in \eqref{QUmid} gives $\widetilde O$ as
\begin{align}\label{Otis}
    \widetilde O=\frac16\eta^{mn} C_{mn}=\frac14\p^m\p^n B_{mn}=\frac1{24}\Box\left(\eta^{mn} B_{mn}\right).
\end{align}
Using these results, the term with the factor $c\ct\p^2\ct\p X^m$, the term with the factor $c\p\ct\p^2\ct$ and the term with the factor $\p^2 c\ct\p\ct$ are all zero. 
Up to now we have
\begin{align}\label{QUmidp}
    &QU=c\p^2c\ct P^m\left(\p^n A_{mn}+F_m+\p_m O\right)+c\ct\p\ct P^m P^n\left(-\frac12\Box A_{mn}+B_{mn}-\p_m F_n\right)\cr 
    &+c\ct\p\ct\p X^m\p X^n\left(-\frac12\Box C_{mn}\right)+c\p^2 c\p\ct\left(\frac12\p^m F_m+K-\frac12\Box O\right)+\ct\p\ct\p^2\ct\left(-\frac12\Box{\widetilde O}\right)=0.
\end{align}

{\small
\bibliographystyle{abe}
\bibliography{mybib}{}
}

\end{document}